\documentclass{PoS}

\usepackage{graphics}
\usepackage{graphicx}
\usepackage{epsfig}
\usepackage{float,amsmath}
\usepackage{amssymb}
\usepackage{times}
\usepackage{fontenc}
\usepackage{mathptmx}

\title{High-energy dileptons from an anisotropic quark-gluon plasma}

\ShortTitle{High-energy dileptons from an anisotropic quark-gluon plasma}

\author{\speaker{Mauricio Martinez}
	\thanks{The author gratefully acknowledges the collaboration and continuous support of Michael Strickland. This work was supported by the Helmholtz Research School and Otto Stern School of the Goethe-Universit\"at Frankfurt am Main.}\\
        Helmholtz Research School and Otto Stern School, Goethe-Universit\"at Frankfurt am Main, Germany\\
        E-mail: \email{guerrero@fias.uni-frankfurt.de}}

\abstract{We calculate leading-order dilepton yields from a quark-gluon plasma which has a time-dependent anisotropy in momentum space. Such anisotropies can arise during the earliest stages of quark-gluon plasma evolution due to the rapid longitudinal expansion of the created matter. A phenomenological model for the proper time dependence of the parton hard momentum scale, $p_{hard}$, and the plasma anisotropy parameter, $\xi$, is proposed. The model describes the transition of the plasma from a 0+1 dimensional collisionally-broadened expansion at early times to a 0+1 dimensional ideal hydrodynamic expansion at late times. We find that high-energy dilepton production is enhanced by pre-equilibrium emission up to 50\% at LHC energies, if one assumes an isotropization/thermalization time of 2 fm/c. Given sufficiently precise experimental data this enhancement could be used to determine the plasma isotropization time experimentally.}

\FullConference{High-pT Physics at LHC -09\\
		 February 4- 4 2009\\
		 Prague, Czech Republic}

\begin{document}

\section{Introduction}
\label{sec:1}

One of the most interesting problems facing the community in relativistic heavy ion collisions is to determine at what time the matter created can be described using hydrodynamics. In this context, at RHIC energies it has been found that for $p_T \lesssim 2$ GeV, the elliptic flow of the matter created is described well by models which assume ideal hydrodynamic behavior starting at very early times $\tau \lesssim$ $1$ fm/c ~\cite{Hirano:2002ds,Tannenbaum:2006ch,Teaney:2000cw,Huovinen:2001cy}. This is not completely understood due to the fact that the estimates from perturbative QCD for the thermalization time of a QGP at RHIC energies range from $2-3$~fm/c~\cite{Baier:2000sb,Xu:2004mz,Strickland:2007fm}. Moreover, recent results from 
conformal relativistic viscous hydrodynamics~\cite{Luzum:2008cw} have shown that these initial estimates for the isotropization/thermalization time of the plasma are not completely reliable due to poor knowledge of the proper initial conditions (CGC versus Glauber), details of plasma hadronization such as the choice of 
the proper freezeout time and the subsequent hadronic cascade, etc. Therefore, additional theoretical 
and experimental inputs are necessary to further constrain this time. It would be nice to have information about the thermalization time and appropiate initial conditions from other independent observables which are different than elliptic flow. One good candidate is high energy dileptons. Due to their large mean free path, lepton pairs can leave the strongly interacting medium after the collision. In this work, we examine the possibility to determine experimentally  the thermalization time of the matter created using high energy dilepton yields as a function of both, mass and transverse momentum. We compute the expected $e^+e^-$ yields resulting from a Pb-Pb collision at LHC full beam energy, $\sqrt{s} = 5.5$ TeV in a pre-equilibrium scenario of a quark-gluon plasma with a time-dependent anisotropy caused by the rapid longitudinal expansion.  We find that at LHC energies, there is an enhancement of the dilepton yields as a function of the transverse momentum when $2< p_T < 8$ GeV if one chooses isotropization/thermalization time of 2 fm/c \cite{Martinez:2008di}.

 This manuscript is organized as follows: In Sec. \ref{sec:1} we calculate the dilepton production rate at leading order using an anisotropic phase space distribution. In Sec. \ref{sec:2} we review the calculation of the dilepton yields from an anisotropic plasma. In Sec. \ref{sec:3} we discuss the main aspects of the interpolating model from collisionally-broadened to ideal hydrodynamical expansion. In Sec. \ref{sec:4} we present the expected medium dilepton yields for LHC energies. Finally, we present our conclusions and give an outlook in the Sec. \ref{sec:5}.
 

\section{Dilepton rate from kinetic theory}
\label{sec:2}
From relativistic kinetic theory, the dilepton production rate 
$dN^{l^+l^-}/d^4Xd^4P\equiv d R^{l^+l^-}/d^4P$ at leading order in the electromagnetic 
coupling, $\alpha$, is calculated as follows: 
\cite{Kapusta:1992uy,Dumitru:1993vz,Strickland:1994rf}:
\begin{equation}
\frac{d R^{l^+l^-}}{d^4P} = \int \frac{d^3{\bf p}_1}{(2\pi)^3}\,\frac{d^3{\bf p}_2}{(2\pi)^3}
			\,f_q({\bf p}_1)\,f_{\bar{q}}({\bf p}_2)\, \it{v}_{q\bar{q}}\,\sigma^{l^+l^-}_{q\bar{q}}\,
			\delta^{(4)}(P-p_1-p_2)
      \; ,
\label{eq:annihilation1}
\end{equation}
where $f_{q,{\bar q}}$ is the phase space distribution function of the 
medium quarks (anti-quarks), $\it{v}_{q\bar{q}}$ is the relative 
velocity between quark and anti-quark and $\sigma^{l^+l^-}_{q\bar{q}}$ 
is the total cross section. Since we will be considering high-energy 
dilepton pairs with center-of-mass energies much greater than the 
dilepton mass we can ignore the finite lepton mass corrections and use 
simply $\sigma^{l^+l^-}_{q\bar{q}} = 4 \pi \alpha^2 / 3 M^2$. In 
addition, we assume that the distribution function of quarks and 
anti-quarks is the same, $f_{\bar q}=f_q$.

We will follow the ansatz by Romatschke and Strickland for the distribution function that describes the pre-equilibrated stage of the plasma \cite{Romatschke:2003ms}. The ansatz consists in consider that any anisotropic distribution function is obtained from an arbitrary isotropic phase space distribution by squeezing ($\xi>0$) or stretching ($\xi<0$) the isotropic distribution function:
\begin{equation}
f_{q,{\bar q}}({\bf p},\xi,p_{\rm hard})=f^{\hspace{0.05cm}iso}_{q,{\bar 
q}}(\sqrt{{\bf p^2}+\xi({\bf p\cdot \hat{n}}){\bf^2}},p_{\rm hard}) \; ,
\label{eq:distansatz}
\end{equation}
where $p_{\rm hard}$ is the hard momentum scale, $\hat{n}$ is the direction of the anisotropy and $\xi >0$ is a parameter that reflects the strength and type of anisotropy. In the isotropic case, $p_{\rm hard}$ is identified with the temperature of the system and $\xi\equiv 0$. In this work we will study the case when the direction of the anisotropy is along the longitudinal (beamline) direction, i.e., $\hat{\bf n} = \hat{e}_z$. This configuration is relevant for earliest stages of the collisions between heavy nuclei.

From Eqs.~(\ref{eq:distansatz}) and~(\ref{eq:annihilation1}) we obtain \footnote{Details of the calculation are presented in Ref.~\cite{Martinez:2008di}.}:
\begin{eqnarray}
\label{anisdilrate}
\frac{d R^{l^+l^-}}{d^4P}&=&
\frac{5\alpha^2}{18\pi^5}\int_{-1}^1d(\cos\theta_{p_1})
\int_{a_+}^{a_-}\frac{dp_1}{\sqrt{\chi}}\,p_1\hspace{0.1cm}f_q\left({\sqrt{\bf p_1^2(1+\xi(\tau)\cos^2\theta_{p_1})}},p_{\rm hard}(\tau,\eta)\right)\nonumber \\
&\times & f_{\bar{q}}\left(\sqrt{{\bf(E-p_1)^2+\xi(\tau)(p_1\cos\theta_{p_1}-P\cos\theta_P)^2}},p_{\rm hard}(\tau,\eta)\right),
\label{scattering}
\end{eqnarray}
with
\begin{eqnarray*}
\chi&=&\,4\,P^2\,p_1^2\,\sin^2\theta_P\,\sin^2\theta_{p_1}-(2p_1(E-P\cos\theta_P\cos\theta_{p_1})-M^2)^2 \; , \\
a_{\pm}&=&\frac{M^2}{2(E-P\cos (\theta_P\pm\theta_{p_1}))} \; .
\end{eqnarray*}
Notice that the dilepton rate does not tell us anything about the space-time evolution of the system and, therefore, it is not enough to make a phenomenological prediction for the expected dilepton yields. In order to make contact with experiments, it is necessary to include the space-time dependence of $p_{\rm hard}$ and $\xi$ and 
then integrate over the space-time volume
\begin{subequations}
\begin{align}
  \frac{dN^{l^+l^-}}{dM^2dy}&=\pi R^2_T\int d^2p_T\int_{\tau_0}^{\tau_f}\int_{-\infty}^{\infty}\frac{dR^{l^+l^-}}{d^4P}\tau d\tau d\eta\hspace{0.2cm}, \label{Mspectrum}\\
        \frac{dN^{l^+l^-}}{d^2p_Tdy}&=\pi R^2_T\int dM^2\int_{\tau_0}^{\tau_f}\int_{-\infty}^{\infty}\frac{dR^{l^+l^-}}{d^4P}\tau d\tau d\eta\hspace{0.2cm},\label{pTspectrum}
\end{align}
\label{spectrumeqs}
\end{subequations}
where $R_T\,=\,1.2\,A^{1/3}$ fm is the radius of the nucleus in the 
transverse plane. These expressions are evaluated in the center-of-mass
(CM) frame while the dilepton production rate is calculated for 
the local rest frame (LR) of the emitting region. Then, the dilepton 
pair energy has to be understood as $E_{LR}=p_T\,\cosh\,(y-\eta)$ in 
the differential dilepton rate $dR_{\rm ann}/d^4P$. 
Substituting Eq.~(\ref{anisdilrate}) into Eqs. (\ref{spectrumeqs}), we obtain the 
dilepton spectrum including the effect of a time-dependent momentum anisotropy. In  Eqs. (\ref{spectrumeqs}) it is assumed that there is only longitudinal expansion of the system. This assumption is well justified since corrections to high energy dileptons coming from transverse expansion or mixed/hadronic phase do not play an important role in the studied kinematic regime \cite{Mauricio:2007vz}. 

Note that we have not included the next-to-leading order (NLO) corrections to the dilepton rate due to the complexity of these contributions for finite $\xi$. These affect dilepton production for isotropic systems for $E/T\lesssim$ 1 \cite{Thoma:1997dk, Arnold:2002ja,Arleo:2004gn,Turbide:2006mc}. In the regions of phase 
space where there are large NLO corrections, we will apply $K$-factors to our results as indicated.


\section{Space-time model}
\label{sec:3}

Previous phenomenological studies of high energy dileptons have assumed that the value of the isotropization time, 
$\tau_{\rm iso}$, is of the same order as the parton formation time $\tau_0$. However, the physical mechanisms which could make such fast isotropization feasible are not completely understood. Recently, a phenomenological model has been proposed where the isotropization time could be larger as a consequence of a previous evolution in a pre-equilibrium dynamics \cite{Martinez:2008di, Mauricio:2007vz}. Before going into the details of the model, we remind the reader of some general relations.

First, the plasma anisotropy parameter $\xi$ is related with the average 
longitudinal momentum ($p_L$) and transverse momentum ($p_T$) of the hard particles through the 
expression:
\begin{equation}
\label{anisoparam}
 \xi=\frac{\langle p_T^2\rangle}{2\langle p_L^2\rangle} - 1 \; .
\end{equation}
Note that this relation is completely general and can be applied in all cases.
It is possible to obtain two important limits from Eq. (\ref{anisoparam}). When we have that  $\langle p_T^2\rangle = 2\langle p_L^2\rangle$, then $\xi = 0$. This is the case when the system is isotropic in momentum-space. Another possibility is when the partons expand freely along the longitudinal axis, i.e., 1d free streaming expansion. Using the free streaming distribution function, it is possible to show that the transverse and longitudinal momentum scales as \cite{Baier:2000sb,Martinez:2008di,Rebhan:2008uj}:
\begin{subequations}
 \begin{align}
  \langle p_T^2 \rangle_{\rm f.s.} &\propto 2 \, T_0^2 \; ,\\
  \langle p_L^2 \rangle_{\rm f.s.} &\propto T_0^2 \frac{\tau_0^2}{\tau^2} \; .
 \end{align}
\end{subequations}
Inserting these expressions into Eq.~(\ref{anisoparam}), one obtains 
$\xi_{f.s.}(\tau) = \tau^2/\tau_0^2-1$.  The free streaming result for 
$\xi$ is the upper-bound on possible momentum-space anisotropies 
developed during 1d expansion by causality. When the system has different kind of interactions, Eq.~(\ref{anisoparam}) for the anisotropy parameter $\xi$ will scale differently as discussed below. 

Second, for a given anisotropic phase space distribution of the 
form specified in Eq.~(\ref{eq:distansatz}), the local energy density 
can be factorized as:
\begin{eqnarray}
\label{energy}
{\cal E}(p_{\rm hard},\xi) &=& \int \frac{d^3{\bf p}}{(2\pi)^3}\hspace{0.1cm}p\hspace{0.1cm}
	f_{\rm iso}(\sqrt{{\bf p^2}+\xi({\bf p\cdot \hat{n}}){\bf^2}},p_{\rm hard}) \; , \\ 
&=& {\cal E}_0(p_{\rm hard}) \, {\cal R}(\xi) \; ,\nonumber
\end{eqnarray}
where ${\cal E}_0$ is the initial local energy density deposited in 
the medium at $\tau_0$ and
\begin{equation}
{\cal R}(\xi) \equiv \frac{1}{2}\Biggl(\frac{1}{1+\xi}+\frac{\arctan\sqrt{\xi}}{\sqrt{\xi}} \Biggr) \; .
\label{calRdef}
\end{equation}
In this work, we will not study explicitly the possibility of 1d free streaming expansion since, in reality, this is a rather extreme assumption which requires that the partons do not interact at all \cite{Martinez:2008di}. 

\subsection{Momentum-space broadening and plasma instabilities effect}

The ratio between the average longitudinal and transverse momentum 
needed to compute $\xi$ using Eq.~(\ref{anisoparam}) is modified from 
the free streaming case if collisions between the partons are taken 
into account. In general, it is a difficult task to obtain the exact form of collisional kernel of the Boltzmann equation plus mean field interactions (Vlasov term). As a first approach, one can start considering the effect 
of elastic collisions in the broadening of the longitudinal momentum of the particles. This was the approach in the original version of the bottom-up scenario \cite{Baier:2000sb}. In the first stage of this scenario, $1\ll Q_s\tau\ll\alpha_s^{3/2}$, initial hard gluons have typical momentum of order $Q_s$ and occupation number of order 
$1/\alpha_s$. Due to the fact that the system is expanding at the speed of light in the longitudinal direction, the density of hard gluons decreases with time, $N_g \sim Q_s^3/(\alpha_s Q_s\tau)$. If there were no interactions this expansion would be equivalent to 1+1 free streaming and the longitudinal momentum $p_L$ would scale like 
$1/\tau$. However, once elastic $2\leftrightarrow 2$ collisions of hard gluons are taken into account, the ratio between the longitudinal momentum $p_L$ and the typical transverse momentum of a hard particle 
$p_T$ decreases as \cite{Baier:2000sb}:
\begin{equation}
\label{ptbroadbottom}
\frac{\langle p_L^2 \rangle}{\langle p_T^2 \rangle} \propto (Q_s\tau)^{-2/3} \; .
\end{equation}
Assuming, as before, isotropy at the formation time, 
$\tau_0=Q_s^{-1}$, this implies that for a collisionally-broadened
expansion, $\xi (\tau)=(\tau/\tau_0)^{2/3}-1$.

Note that in order to obtain the relation given by Eq. (\ref{ptbroadbottom}) it is implicitly assumed that that the elastic cross-section is screened at long distances by an isotropic real-valued Debye mass \cite{Baier:2000sb}. But this is not the case of an anisotropic plasma, since the Debye mass can become complex due to the chromo-Weibel instability \cite{Romatschke:2003ms,Rebhan:2008uj,Mrowczynski:2000ed,Arnold:2003rq, Romatschke:2004jh}. Such negative eigenvalues indicate instabilities, which result in exponential growth of chromo-electric and magnetic fields, $E^a$ and $B^a$, respectively. These fields deflect the particles and how much deflection occurs will depend on the amplitude and domain size of the induced chromofields. Currently, the precise parametric relation between the plasma anisotropy parameter and the amplitude and domain size of the chromofields is not known from first principles.

If one would like to include the momentum-space broadening and the 
impact of the plasma instabilities, this can be achieved by considering the temporal dependence of $\xi (\tau)$ as:
\begin{equation}
\xi(\tau,\delta) = \left( \frac{\tau}{\tau_0} \right)^\delta - 1 \; .
\label{broadenedxi}
\end{equation}
The exponent $\delta$ contains the physical information about the 
particular type of momentum-space broadening which occurs due to plasma
interactions. Two limiting cases for this exponent are 
the ideal hydrodynamic and free streaming expansion. In the 1+1 
hydrodynamical limit, $\delta \equiv 0$ and then $\xi \rightarrow 0$. 
For  $\delta \equiv 2$, one recovers the 1+1 dimensional free 
streaming case, $\xi \rightarrow \xi_{\rm f.s.} = (\tau /\tau_0)^2-1$. 
For  $0 < \delta < 2$, one obtains the proper-time dependence of the energy density and temperature by substituting 
(\ref{broadenedxi}) into the general expression for the factorized energy density (\ref{energy}) to obtain 
${\cal E}(\tau,\delta) = {\cal E}_0(p_{\rm hard}) \, {\cal R}(\xi(\tau,\delta))$. 

Different values of $\delta$ arise dynamically from the different 
processes contributing to parton isotropization. For example, elastic
collisional-broadening results in Eq.~(\ref{ptbroadbottom}) and hence 
$\delta=2/3$.  Recently, some authors have considered the values of 
$\delta$ resulting from processes associated with the chromo-Weibel 
instability presented at the earliest times after the initial nuclear 
impact \cite{Bodeker:2005nv,Arnold:2005qs,Arnold:2007cg}:
\begin{equation}
\label{ptbroadproposals1}
\frac{\langle p_L^2 \rangle}{\langle p_T^2 \rangle}\sim(Q_s\tau)^{-\frac{1}{2}\bigl(\frac{1}{1+\nu}\bigl)} \; ,
\end{equation}
where
\begin{equation}
 \label{ptbroadproposals2}
\nu=\left\{ \begin{aligned}
0 \hspace{0.2cm}&\text{Ref.\cite{Bodeker:2005nv} \; ,}\\
1 \hspace{0.2cm}&\text{Ref.\cite{Arnold:2005qs} \; ,}\\
2 \hspace{0.2cm}&\text{Nielsen-Olesen limit, Ref.\cite{Arnold:2007cg}} \;.
          \end{aligned}
	  \right.
\end{equation}
Summarizing, the coefficient $\delta$ in Eq.~(\ref{broadenedxi}) takes 
on the following values
\begin{equation} 
\label{anisomodels}
\delta=\left\{ 
\begin{aligned}
2 \hspace{0.2cm}&\text{Free streaming expansion} \; ,\\
2/3 \hspace{0.2cm}&\text{Collisional-Broadening, Ref.\cite{Baier:2000sb}\; ,}\\
1/2 \hspace{0.2cm}&\text{Ref.\cite{Bodeker:2005nv}}\; , \\
1/4 \hspace{0.2cm}&\text{Ref.\cite{Arnold:2005qs}\; ,}\\
1/6 \hspace{0.2cm}&\text{Nielsen-Olesen limit, Ref.\cite{Arnold:2007cg}}\; ,\\
0\hspace{0.2cm}&\text{Hydrodynamic expansion} \; .
\end{aligned}
\right.
\end{equation}
%

\begin{figure*}[h!]
\begin{center}
 \includegraphics[scale=1.2]{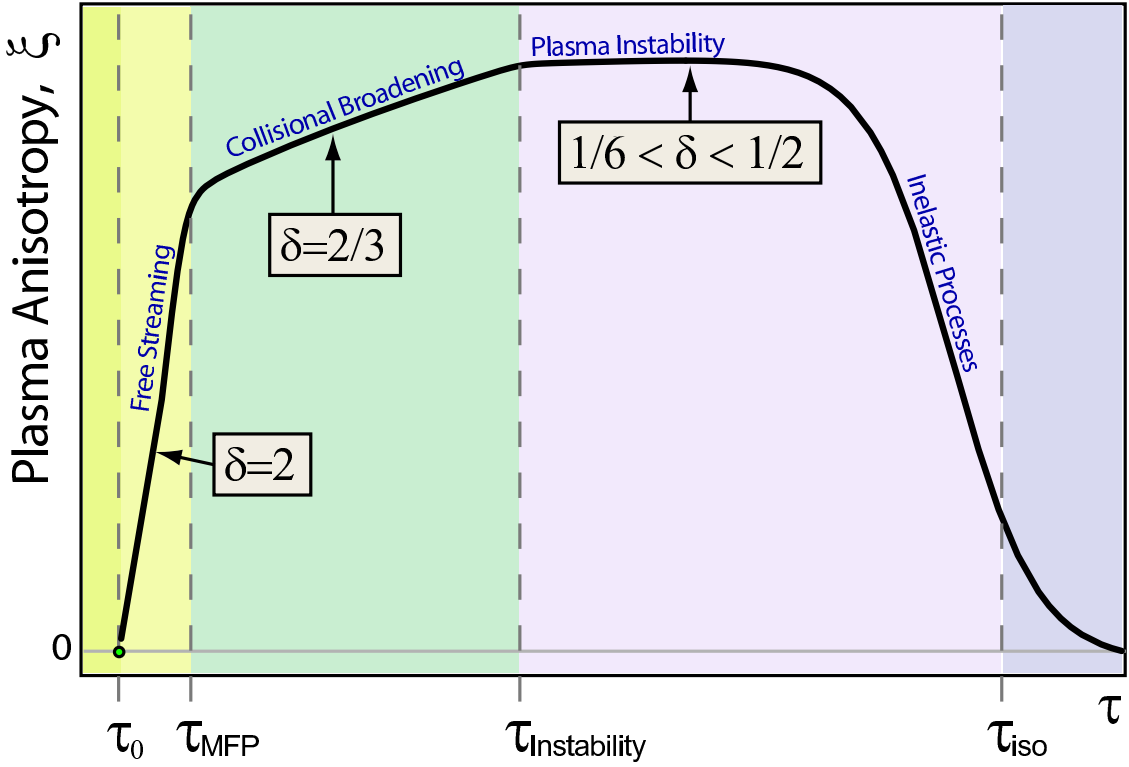}
\end{center}
\vspace{-2mm}
\caption{Sketch of the time dependence of the plasma anisotropy indicating the 
various time scales and processes taking place.  Here $\tau_{\rm MFP}$ 
is the mean time between elastic collisions (mean-free time) and $\tau_{\rm 
Instability}$ is the time at which plasma-instability induced soft 
modes have grown large enough to affect hard particle dynamics.}
\label{fig:1}
\end{figure*}

In Fig.~\ref{fig:1} we sketch the time-dependence of the plasma 
anisotropy parameter indicating the time scales at which the various 
processes become important.  At times shorter than the mean time 
between successive elastic scatterings, $\tau_{\rm MFP}$, the system 
will undergo 1+1 dimensional free streaming with $\delta=2$. For times 
long compared to $\tau_{\rm MFP}$ but short compared to $\tau_{\rm 
Instability}$ the plasma anisotropy will grow with the 
collisionally-broadened exponent of $\delta=2/3$. Here 
$\tau_{\rm Instability}$ is the time at which instability-induced soft gauge 
fields begin to influence the hard-particles' motion. When $\tau_{\rm 
Instability} < \tau < \tau_{\rm iso}$ the plasma anisotropy grows with 
the slower exponent of $\delta = 1/6 \ldots 1/2$ due to the bending of 
particle trajectories in the induced soft-field background.  At times 
large compared to $\tau_{\rm Instability}$ inelastic processes are 
expected to drive the system back to isotropy \cite{Baier:2000sb}. We 
note here that for small $\xi$ and realistic couplings it has been 
shown \cite{Schenke:2006xu} that one cannot ignore the effect of 
collisional-broadening of the distribution functions and that this may 
completely eliminate unstable modes from the spectrum.

From this sketch, it may be possible to try to construct a detailed model which includes all of the various time scales and study the dependence of the process under consideration on each. However, there are theoretical uncertainties in each of these time scales and their dependences on experimental conditions. Because of this, we will construct a phenomenological model which smoothly interpolates the coefficient $\delta$ from the 1d collisionally-broadened expansion to 1d hydrodynamical expansion, i.e., $2/3\geq \delta \geq 0$.

In the model we introduce a transition width, $\gamma^{-1}$, which indicates the smoothness of the transition from the initial value of $\delta =2/3$ to $\delta=0$ at $\tau \sim \tau_{\rm iso}$. Note that by using such a smooth interpolation one can achieve a reasonable phenomenological description of the transition from non-equilibrium to equilibrium dynamics which should hopefully capture the essence of the physics. The collisionally-broadened interpolating model provides us a realistic estimate of the effect of plasma anisotropies.
\subsection{Interpolating model for collisionally broadened expansion}

In order to construct an interpolating model between 
collisionally-broadened and hydrodynamical expansion, we introduce 
the smeared step function:
\begin{equation}
\lambda(\tau,\tau_{\rm iso},\gamma) \equiv \frac{1}{2} \left({\rm 
tanh}\left[\frac{\gamma (\tau-\tau_{\rm iso})}{\tau_{\rm iso}} \right]+1\right) \; ,
\end{equation}
where $\gamma^{-1}$ sets the width of the transition between non-equilibrium and hydrodynamical evolution in units of $\tau_{\rm iso}$. In the limit when $\tau \ll \tau_{\rm iso}$, we have $\lambda \rightarrow 0$ and when $\tau \gg \tau_{\rm iso}$ we have $\lambda \rightarrow 1$. Physically, the energy density ${\cal E}$ should be continuous as we change from the initial non-equilibrium value of $\delta$ to the final isotropic $\delta=0$ value appropriate for ideal hydrodynamic expansion.  Once the energy density is specified, this gives us the time dependence of the hard momentum scale. We find that for general $\delta$ this can be accomplished with the following model for fixed final multiplicity \cite{Martinez:2008di}:
\begin{subequations}
\label{eq:modelEQs}
\begin{align}
\xi(\tau,\delta) &= \left(\tau/\tau_0\right)^{\delta(1-\lambda(\tau))} - 1 \; , \label{xidependence}\\
{\cal E}(\tau, \eta) &= {\cal E}_0 \; {\cal R}\left(\xi\right) \; \bar{\cal U}^{4/3}(\tau) \; ,\\
p_{\rm hard}(\tau, \eta) &= T_0 \; \bar{\cal U}^{1/3}(\tau) \; ,
\label{pdependence}
\end{align}
\end{subequations}
with ${\cal R}(\xi)$ defined in Eq.~(\ref{calRdef}) and for fixed final multiplicity we have:
\begin{subequations}
\begin{eqnarray}
\bar{\cal U}(\tau) &\equiv& {\cal U}(\tau) \, / \, {\cal U}(\tau_{\rm iso}^+) \; , \\
{\cal U}(\tau) &\equiv& \left[{\cal R}\!\left(\left(\tau_{\rm iso}/\tau_0\right)^\delta- 
1\right)\right]^{3\lambda(\tau)/4} \left(\frac{\tau_{\rm iso}}{\tau}\right)^{1 - \delta\left(1-\lambda(\tau)\right)/2} \; , \nonumber \\
{\cal U}(\tau_{\rm iso}^+) &\equiv& \lim_{\tau\rightarrow\tau_{\rm iso}^+} {\cal U}(\tau) 
  = \left[{\cal R}\!\left(\left(\tau_{\rm iso}/\tau_0\right)^\delta-1\right)\right]^{3/4} 
	  \left(\frac{\tau_{\rm iso}}{\tau_0}\right) \; .
\end{eqnarray}
\label{Udeff}
\end{subequations}
and $\delta = 2/3$ for the case of 1d collisionally broadened expansion
interpolating to 1d ideal hydrodynamic expansion.

\begin{figure*}[h]
\begin{center}
 \includegraphics[width=15cm, height=5.7cm]{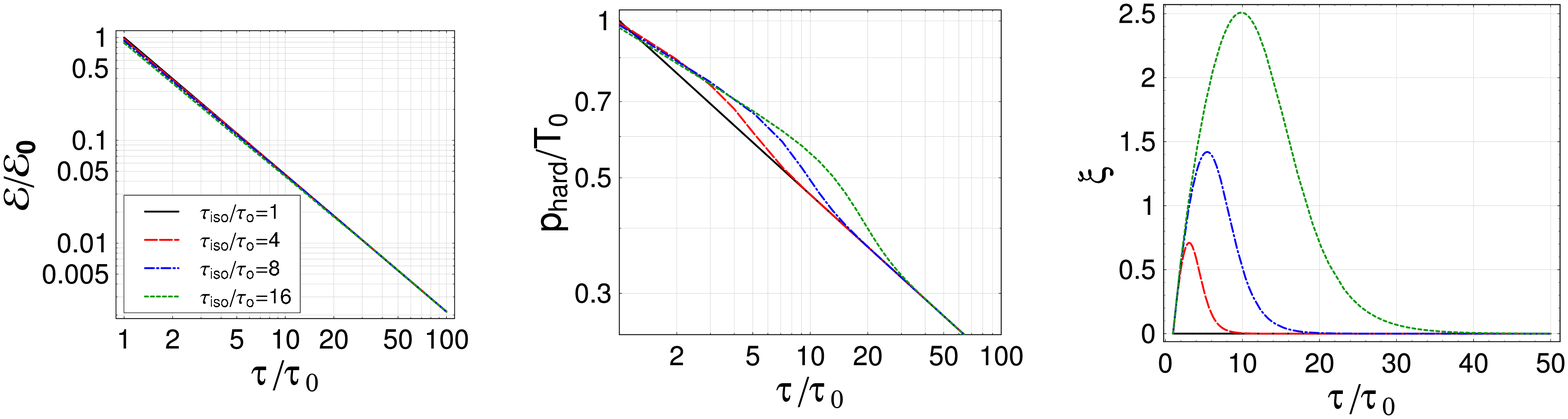}%
\end{center}
\vspace{-2mm}
\caption{Temporal evolution using our fixed final multiplicity interpolating models
for the energy density (left column), hard momentum 
scale (middle column), and anisotropy parameter (right column) for 
four different isotropization times $\tau_{\rm iso} \in \{1,4,6,18\} 
\, \tau_0$. To convert to physical scales use $\tau_0 \sim 0.3$ fm/c for RHIC and $\tau_0 \sim 0.1$ fm/c for LHC.}
\label{fig:2}
\end{figure*}


In Fig.~\ref{fig:2}, the temporal evolution of the energy density, hard momentum scale and the anisotropy 
parameter $\xi(\tau)$ is plotted using Eq.~(\ref{xidependence}). As can be seen from the figure for fixed final multiplicity at $\tau = \tau_0$, the initial value of $p_{hard}$ is reduced for finite values of $\tau_{\rm iso}$ but once the system looks isotropic in momentum-space around $\tau\gtrsim\tau_{\rm iso}$, $p_{\rm hard}$ is the same, independent of its initial value \cite{Martinez:2008di}. 


\section{Results}
\label{sec:4}
We calculate predicted $e^+e^-$ yields as a function of invariant mass and transverse momentum along with predicted yields from other sources for LHC energies. For comparison with previous works we take $\tau_0 = 0.088$ fm/c, $T_0 = 845$ MeV, $T_c = 160$ MeV, and $R_T = 7.1$ fm~\cite{Turbide:2006mc}. Here we assume that when the system reaches $T_c$ all medium emission stops. 

\begin{figure*}[h]
\begin{center}
\includegraphics[scale=0.57]{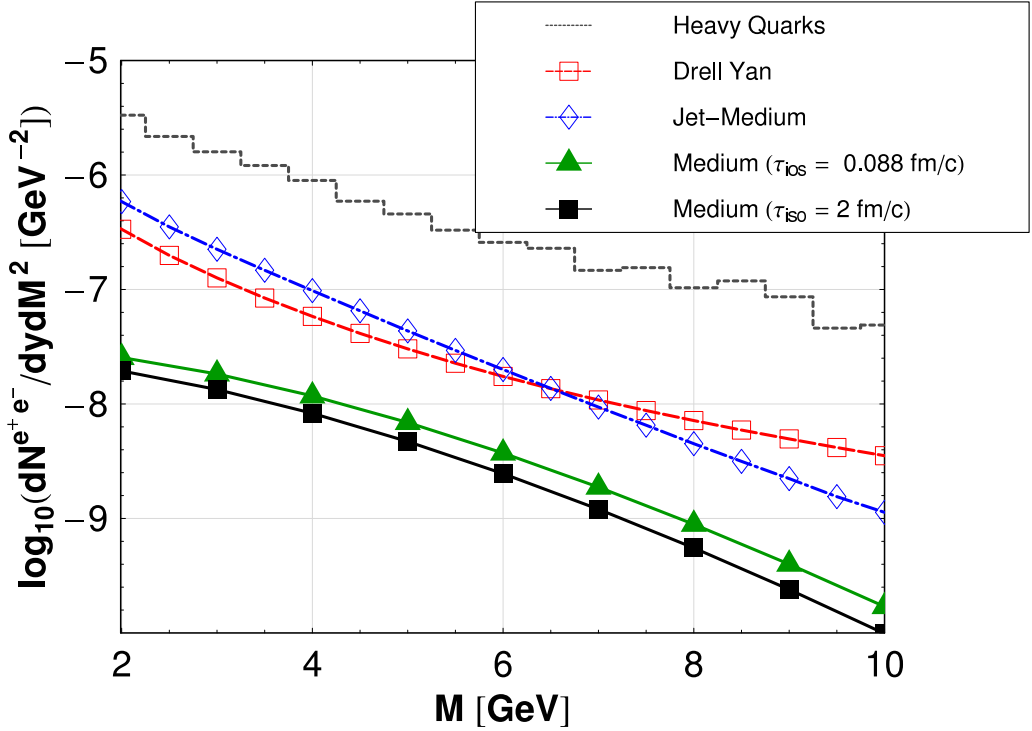}
\end{center}
\vspace{-2mm}
\caption{Collisionally-broadened interpolating model dilepton yields as a function of invariant mass in central 
Pb+Pb collisions at the LHC, with a 
cut $p_T\,\geq$ 8 GeV and and rapidity $y$=0. For medium dileptons we use $\gamma$=2 and 
$\tau_{\rm iso}$ = 0.088 fm/c for LHC energies and fixed final multiplicity. A $K$-factor of 1.5 was applied to account for NLO corrections. Dilepton yields from Drell Yan, Heavy Quarks, Jet-Thermal and Jet-Fragmentation were obtained from Ref.~\cite{Turbide:2006mc}.}
\label{dilmassmultbroad}
\end{figure*}

\begin{figure*}[h]
\begin{center}
\includegraphics[scale=0.57]{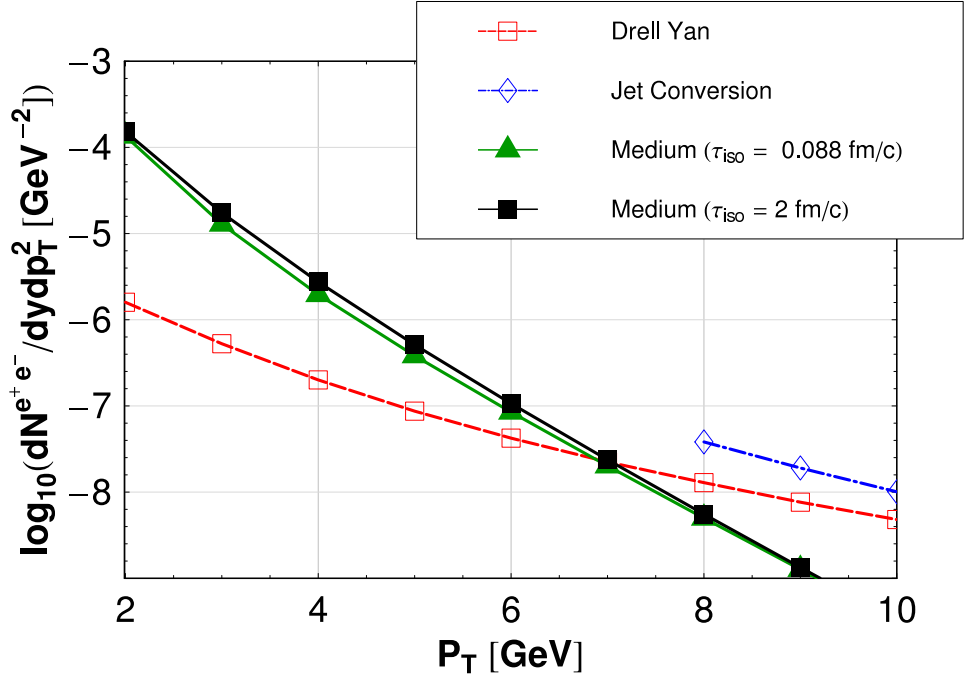}
\end{center}
\vspace{-2mm}
\caption{Collisionally-broadened interpolating model dilepton yields as a function of transverse momentum in 
central Pb+Pb collisions at the LHC, with a cut $0.5\,\leq\,M\,\leq\,1$ GeV and rapidity $y$=0. 
For medium dileptons we use $\gamma$=2 and $\tau_{\rm iso}$ = 0.088 fm/c for LHC energies 
and fixed final multiplicity. A $K$-factor of 6 was applied to account for NLO corrections. Dilepton yields from Drell Yan, 
Jet-Thermal and Jet-Fragmentation were obtained from Ref.~\cite{Turbide:2006mc}.}
\label{dilptmultbroad}
\end{figure*}

In Fig.~\ref{dilmassmultbroad} we show our predicted dilepton mass spectrum for LHC energies using the time dependence of the energy density, the hard momentum scale and the anisotropy parameter given by Eqns.~(\ref{eq:modelEQs}) with $\delta=2/3$. As can be seen from Fig.~\ref{dilmassmultbroad} there is a difference of the medium dilepton yield when varying the assumed plasma isotropization time from $0.088$ fm/c to 2 fm/c. However, the prediction is up to one order of magnitude below the other contributions of dilepton yields (Drell-Yan, jet-thermal, and jet-fragmentation). This coupled with the large background coming from semileptonic heavy quarks decays would make it extremely difficult for experimentalists to extract a clean medium dilepton signal from the invariant mass spectrum. For this reason it does not look very promising to determine plasma initial conditions from the dilepton invariant mass spectrum. Nevertheless, the situation looks better for the dilepton spectrum as a function of the transverse momentum as it is shown in Fig.~\ref{dilptmultbroad}. In this spectrum, there is an enhancemment of medium dilepton yield for $2 < p_T < 8$ GeV compared with the other sources of dileptons (Drell Yann and Jet conversion). Therefore, this observable offers the oportunity to have a cleanest way to determine plasma initial conditions. 

\begin{figure*}[h]
\begin{center}
\includegraphics[scale=0.57]{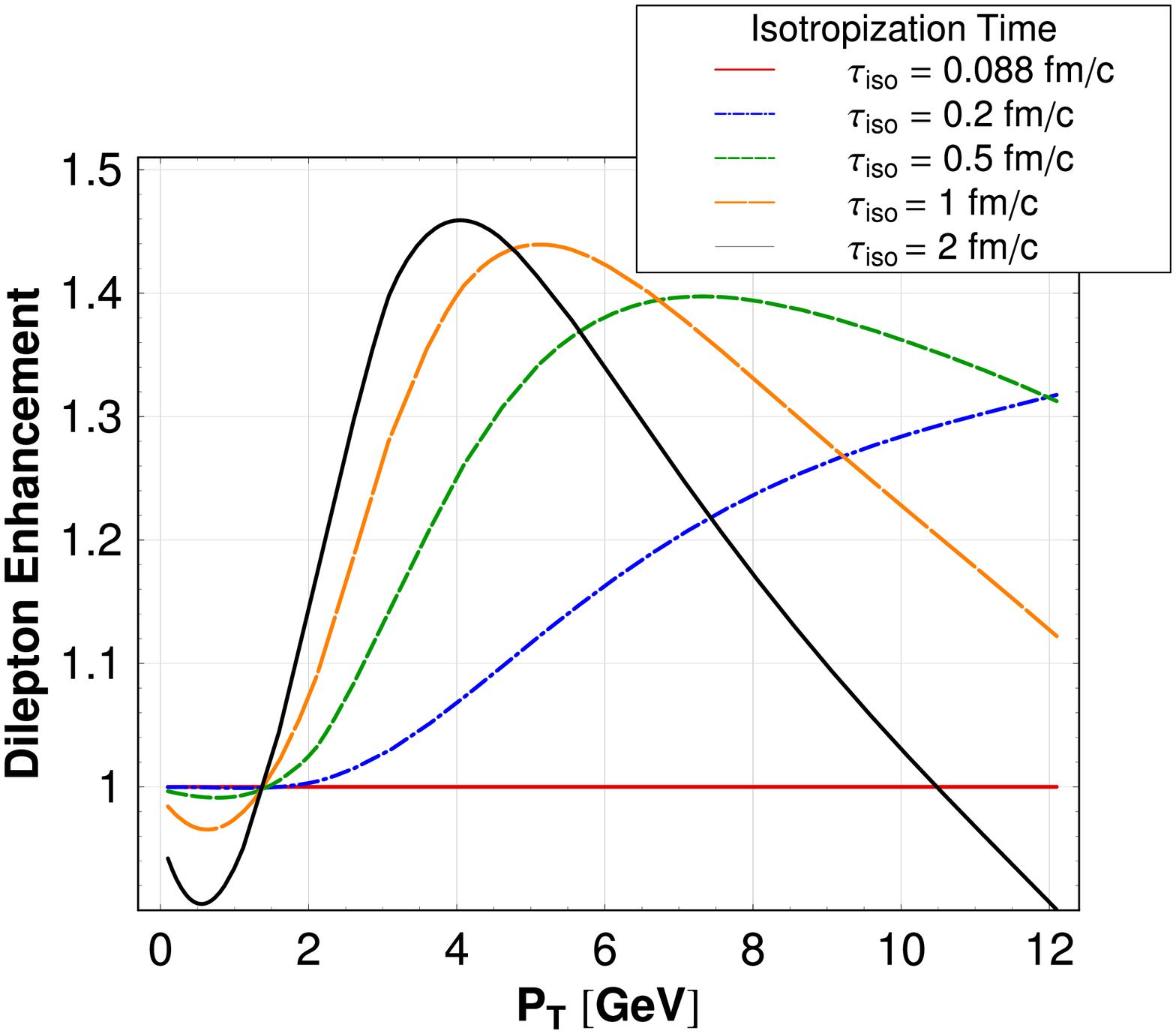}
\end{center}
\vspace{-2mm}
\caption{
Dilepton enhancement, $\phi$ as defined in Eq.~(4.1)
resulting from our collisionally-broadened interpolating model 
with fixed final multiplicity and  $\gamma$ = 2.  
The invariant mass cut used was $0.5\,<\,M\,<\,1$ GeV, rapidity $y$=0 and $\tau_{0}=$ 0.088 fm/c for LHC energies. 
Lines show expected pre-equilibrium dilepton
enhancements for different values of the assumed plasma isotropization time, $\tau_{\rm iso}$.
}
\label{enhance1-mult}
\end{figure*}

In order to quantify the effect of time-dependent pre-equilibrium 
emissions we define the ``dilepton enhancement'', $\phi(\tau_{\rm 
iso})$, as the ratio of the dilepton yield obtained with an isotropization time 
of $\tau_{\rm iso}$ to that obtained from an instantaneously 
thermalized plasma undergoing only 1+1 hydrodynamical expansion, ie. 
$\tau_{\rm iso}=\tau_0$.
\begin{equation}
\phi(\tau_{\rm iso}) \equiv \left. \left( \dfrac{dN^{e^+e^-}(\tau_{\rm iso})}{dy dp_T^2} \right) \right/
\left( \dfrac{dN^{e^+e^-}(\tau_{\rm iso}=\tau_0)}{dy dp_T^2} \right).
\label{dileptonenhancement}
\end{equation}
This ratio measures how large the effect of early-time momentum 
anisotropies are on medium dilepton production. In the case of 
instantaneous isotropization, $\Phi(\tau_{\rm iso})$ is unity, and for 
$\tau_{\rm iso} > \tau_0$ any deviation from unity indicates a 
modification of medium dilepton production due to pre-equilibrium 
emissions. 

In Fig.~\ref{enhance1-mult} we show the dilepton enhancement, $\phi$, as 
function of transverse momentum for $\tau_{\rm iso} = 2$ fm/c at LHC energies. The invariant mass cut is 
the same as in Fig.~\ref{dilptmultbroad} ($0.5\,\leq\,M\,\leq\,1$ 
GeV).  As can be seen from Fig.~\ref{enhance1-mult}, there is a rapid increase in $\phi$ 
1 and 4 GeV at LHC energies. Moreover, from this figure both sharp and smooth 
transitions from early-time collisionally-broadened expansion to ideal 
hydrodynamic expansion result in a 30-50\% at LHC energies. The effect of reducing $\tau_{\rm iso}$ is to shift the peak
in $\phi$ to larger $p_T$ while at the same time reducing the overall amplitude of the peak. Therefore, in order to see the difference between an instantaneously thermalized QGP with $\tau_{\rm iso} = \tau_0$ and one with a later thermalization time requires determining the medium dilepton spectra between 2 < $p_T$ < 8 GeV at LHC with high precision so that one could measure the less than 50\% variation resulting from pre-equilibrium emissions.
Finally, we point out that it is possible to take other cuts (invariant mass and/or transverse momentum).
This could be coupled with fits to experimental data, allowing one to fix $\tau_{\rm iso}$ and $\gamma$ via a ``multiresolution'' analysis.


\section{Conclusions}
\label{sec:5}

In this work we have introduced a phenomenological model that takes into account early-time momentum-space anisotropies in the rapidity dependence of high-energy dilepton production. To do this we have modeled the temporal evolution of the plasma anisotropy parameter $\xi$ and the hard momentum scale $p_{\rm hard}$.  Using these models 
we integrated the leading order rate for dilepton production in an anisotropic plasma over our modeled space-time evolution. Based on our numerical results for the variation of dilepton yields with the assumed values of $\tau_{\rm iso}$, we find that the best opportunity to determine information about the plasma isotropization time is by 
analyzing the high transverse momentum (2 $< p_T <$ 8 GeV at LHC) dilepton spectra using relatively low pair 
invariant mass cuts ($M \lesssim 2$ GeV).  Based on these $p_T$ spectra we introduced the ``dilepton enhancement'' factor $\phi(\tau_{\rm iso})$ which measures the ratio of yields obtained from a plasma which isotropizes at $\tau_{\rm iso}$ to one which isotropizes at the formation time, $\tau_0$. We find that there is a 30-50\% enhancement in the high-transverse momentum dileptons at LHC energies when one assumes an isotropization time of $\tau_{\rm iso} = 2$ fm/c.  The amplitude of the enhancement and position of the peak in the enhancement function, $\phi$, varies with the assumed value of $\tau_{\rm iso}$ which, given sufficiently precise data, would provide a way to determine the plasma isotropization time experimentally. Another interesting possibility to constrain the value to $\tau_{\rm iso}$ using dileptons is studying the forward region, where the effect of early-time anisotropies is expected to be maximal \cite{Martinez:2008mc}.
 
As an extension of the interpolating model presented in this work, one can consider the case where, instead of having at late times 1+1 ideal hydrodynamical expansion, the system is described during that stage by 1+1 viscous hydrodynamical expansion. This is possible since one can relate the anisotropy parameter, $\xi$, with the anisotropy in momentum-space of the fluid generated by the difference of the pressures along the longitudinal and transverse plane due to viscous corrections \cite{Martinez:2009mf}. 

An uncertainty of our treatment comes from our implicit assumption of chemical equilibrium. If the system is not in chemical equilibrium (too many gluons and/or too few quarks) early time quark chemical potentials, or fugacities, will affect the production of lepton pairs \cite{Dumitru:1993vz,Strickland:1994rf}.  However, to leading order, 
the quark and gluon fugacities will cancel between numerator and denominator in the dilepton suppresion factor, $\phi(\tau_{\rm iso})$ \cite{Strickland:1994rf}.  We, therefore, expect that to good approximation one can factorize the effects of momentum space anisotropies and chemical non-equilibrium.

We note in closing that within this model it is possible to assess the phenomenological consequences of momentum-space anisotropies on other observables which are sensitive to early-time stages of the QGP, e.g. photon production, heavy-quark transport, jet-medium induced electromagnetic and gluonic radiation, etc.

\end{document}